\documentclass[preprint]{iucr}              

     \journalcode{M}              

\begin{document}                  



\title{Unsupervised learning approaches to characterize heterogeneous samples using X-ray single particle imaging}

\author[a,b]{Yulong}{Zhuang}
\author[c]{Salah}{Awel}
\author[c]{Anton}{Barty}
\author[d]{Richard}{Bean}
\author[d]{Johan}{Bielecki}
\author[d]{Martin}{Bergemann}
\author[e]{Benedikt J.}{Daurer}
\author[f]{Tomas}{Ekeberg}
\author[c]{Armando D.}{Estillore}
\author[a,b,d,r]{Hans}{Fangohr}
\author[d]{Klaus}{Giewekemeyer}
\author[g]{Mark S.}{Hunter}
\author[d]{Mikhail}{Karnevskiy}
\author[h]{Richard A.}{Kirian}
\author[d]{Henry}{Kirkwood}
\author[d]{Yoonhee}{Kim}
\author[d]{Jayanath}{Koliyadu}
\author[i,j]{Holger}{Lange}
\author[d]{Romain}{Letrun}
\author[c,i,k]{Jannik}{L{\"u}bke}
\author[a,b]{Abhishek}{Mall}
\author[d]{Thomas}{Michelat}
\author[l]{Andrew J.}{Morgan}
\author[c,k]{Nils}{Roth}
\author[c]{Amit K.}{Samanta}
\author[d]{Tokushi}{Sato}
\author[e]{Zhou}{Shen}
\author[c]{Marcin}{Sikorski}
\author[i,s]{Florian}{Schulz}
\author[h]{John C.~H.}{Spence}
\author[c,d]{Patrik}{Vagovic}
\author[a,b,i]{Tamme}{Wollweber}
\author[c,k]{Lena}{Worbs}
\author[a,c,i]{P. Lourdu}{Xavier}
\author[c]{Oleksandr}{Yefanov}
\author[f,m]{Filipe R.~N.~C.}{Maia}
\author[c,i,n]{Daniel A.}{Horke}
\author[c,i,k,o]{Jochen}{K{\"u}pper}
\author[e,p]{N. Duane}{Loh}
\author[d,q]{Adrian P.}{Mancuso}
\author[c,i,k]{Henry N.}{Chapman}
\cauthor[a,b,i]{Kartik}{Ayyer}{kartik.ayyer@mpsd.mpg.de}

\aff[a]{Max Planck Institute for the Structure and Dynamics of Matter, 22761 Hamburg, Germany}
\aff[b]{Center for Free-Electron Laser Science, 22761 Hamburg, Germany}
\aff[c]{Center for Free-Electron Laser Science, DESY, 22607 Hamburg, Germany}
\aff[d]{European XFEL, 22869 Schenefeld, Germany}
\aff[e]{Center for BioImaging Sciences, National University of Singapore, Singapore 117557}
\aff[f]{Dept. of Cell and Molecular Biology, Uppsala University, 75124 Uppsala, Sweden}
\aff[g]{Linac Coherent Light Source, SLAC National Accelerator Laboratory, Menlo Park, 94025, USA}
\aff[h]{Department of Physics, Arizona State University, Tempe, AZ 85287, USA}
\aff[i]{The Hamburg Center for Ultrafast Imaging, Universit{\"a}t Hamburg, 22761 Hamburg, Germany}
\aff[j]{Institute of Physical Chemistry, Universit{\"a}t Hamburg, 20146 Hamburg, Germany}
\aff[k]{Department of Physics, Universit{\"a}t Hamburg, 22761 Hamburg, Germany}
\aff[l]{Univ. of Melbourne, Physics, Victoria, 3010, Australia}
\aff[m]{NERSC, Lawrence Berkeley National Laboratory, Berkeley, CA 94720, USA}
\aff[n]{Institute for Molecules and Materials, Radboud University, 6525 AJ Nijmegen, Netherlands}
\aff[o]{Department of Chemistry, Universit{\"a}t Hamburg, 20146 Hamburg, Germany}
\aff[p]{Department of Physics, National University of Singapore, Singapore 117551}
\aff[q]{Department of Chemistry and Physics, La Trobe Institute for Molecular Science, La Trobe University, Melbourne, Victoria 3086, Australia}
\aff[r]{University of Southampton, SO17 1BJ, Southampton, United Kingdom}
\aff[s]{Institute of Nanostructure and Solid State Physics, University of Hamburg, Luruper Chaussee 149, 22761 Hamburg, Germany}

\shortauthor{Zhuang et al.}

\maketitle                        

\begin{synopsis}
Two unsupervised machine learning approaches are presented to characterize structural variations in nanoparticle ensembles measured using X-ray single particle imaging. The algorithms are applied to an experimental dataset where both discrete structural classes as well as continuous deformations caused by X-ray induced melting are classified.
\end{synopsis}

\begin{abstract}
One of the outstanding analytical problems in X-ray single particle imaging (SPI) is the classification of structural heterogeneity, which is especially difficult given the low signal-to-noise ratios of individual patterns and that even identical objects can yield patterns that vary greatly when orientation is taken into consideration. We propose two methods which explicitly account for this orientation-induced variation and can robustly determine the structural landscape of a sample ensemble. The first, termed common-line principal component analysis (PCA) provides a rough classification which is essentially parameter-free and can be run automatically on any SPI dataset. The second method, utilizing variation auto-encoders (VAEs) can generate 3D structures of the objects at any point in the structural landscape. We implement both these methods in combination with the noise-tolerant expand-maximize-compress (EMC) algorithm and demonstrate its utility by applying it to an experimental dataset from gold nanoparticles with only a few thousand photons per pattern and recover both discrete structural classes as well as continuous deformations. These developments diverge from previous approaches of extracting reproducible subsets of patterns from a dataset and open up the possibility to move beyond studying homogeneous sample sets and study open questions on topics such as nanocrystal growth and dynamics as well as phase transitions which have not been externally triggered.



\end{abstract}

 
\section{Introduction}

X-ray single particle imaging (SPI) is a method to reconstruct 3D structures of isolated nanoscale objects by collecting a large number of diffraction patterns using bright X-ray pulses. The diffraction patterns sample the object's three-dimensional Fourier transform along randomly oriented spherical slices, which enables a Fourier synthesis of the 3D structure as long as the orientations can be determined~\cite{Neutze:2000}. However, the X-ray pulses are bright enough to destroy the sample after each shot, and so each pattern is collected from a different particle. If the particles are reproducible up to the resolution of interest, then determination of 3D structures is fairly straightforward, involving the determination of the orientation and incident X-ray fluence for each shot. Various algorithms have been proposed and implemented for this purpose including the expand–maximize–compress (EMC) algorithm~\cite{Loh:2009}, which has been used for a number of experimental demonstrations~\cite{Ekeberg2015,Rose2018,Shi2018,Ayyer:2019,Ayyer:2021}. Other methods utilizing intensity correlations have also been experimentally demonstrated~\cite{Kurta2017,vonArdenne2018}.

However, one of the challenges in analyzing a serial dataset like that produced in an SPI experiment is the proper classification of patterns in terms of their structures. This is a necessary step in order to obtain a high resolution structure since the underlying assumption of the above mentioned algorithms is that the particles are identical, which is never true in practice. On the other hand, the framework used for this classification can be used to study datasets where the heterogeneity is not just a drawback, uncovering the landscape of structural variations in the sample. This requires the detection of discrete classes of object shapes representing contaminants, aggregates etc. and even the detection of continuous deformations depending on the scientific problem being studied.



In past years, different machine learning algorithms have been developed for use in this classification of structural variability. These algorithms are usually applied to the patterns themselves using methods including spectral clustering~\cite{Yoon:2011}, support vector machines~\cite{Bobkov:2015}, and convolutional neural networks~\cite{Zimmermann:2019,Ignatenko:2021}. 

In this work, we take an unsupervised learning approach to analyze an experimental dataset of more than 2.4 million diffraction patterns of nominally 42~nm cubic gold nanoparticles collected at the European XFEL~\cite{Ayyer:2021}. The goal is to classify the entire structural ensemble, including separating discrete contaminants as well as to study any continuous structural variations that may be present. We discuss two approaches, the first of which produces an embedding of the diffraction patterns based on their 3D structures. This approach can be applied without modification to most SPI datasets. The second is a generative method using variational auto-encoders (VAEs) which enables us to visualize the 3D structure of the particle at any point along its landscape. Both methods are applied to the dataset to study a continuous, XFEL-induced deformation of the cubic nanoparticles to spheres.


\section{Results}

\subsection{Experiment and dataset information}
The dataset discussed in the rest of this work was collected as part of the experiment described in \citeasnoun{Ayyer:2021}, which we review in brief: The SPI experiment was performed with the MHz-rate European XFEL~\cite{Decking:2020} at the SPB/SFX (single particles, biomolecules, and clusters/serial femtosecond crystallography) instrument~\cite{Mancuso:2019} with 6 keV photons in pulses with an average energy of 2.5 mJ ($2.6 \times 10^{12}$ photons) measured upstream of the focusing optics. The Adaptive Gain Integrating Pixel Detector (AGIPD)~\cite{Allahgholi:2019}, was placed 705 mm downstream of the interaction region to collect each diffraction pattern up to a scattering angle of 8.3$^{\circ}$. In this analysis, we will use data from the central region of the detector up to 1.8$^{\circ}$.

The samples were nominally gold cubes with edge lengths 42~nm and were injected into the X-ray beam via an electrospray aerosolization aerodynamic-lens-stack sample delivery system. For this sample, 34,197,950 frames were recorded of which 2,451,068 were judged to contain sample diffraction (hit ratio $\sim$ $7.17\%$). Part of the dataset was collected with the European XFEL running at an intra-train repetition rate of 1.1 MHz, during which a high fraction of the diffraction patterns originated from spherical particles. This was found to be due to the pre-exposure of particles in the wings of the previous XFEL pulse in the train, which seemed to lead to melting. This effect disappeared when reducing the repetition rate to 0.5 MHz. As a result, the sample contains diffraction patterns of cubes, spheres and potentially cubes at different stages of melting/softening.  

\begin{figure}
\includegraphics[width=0.95\textwidth]{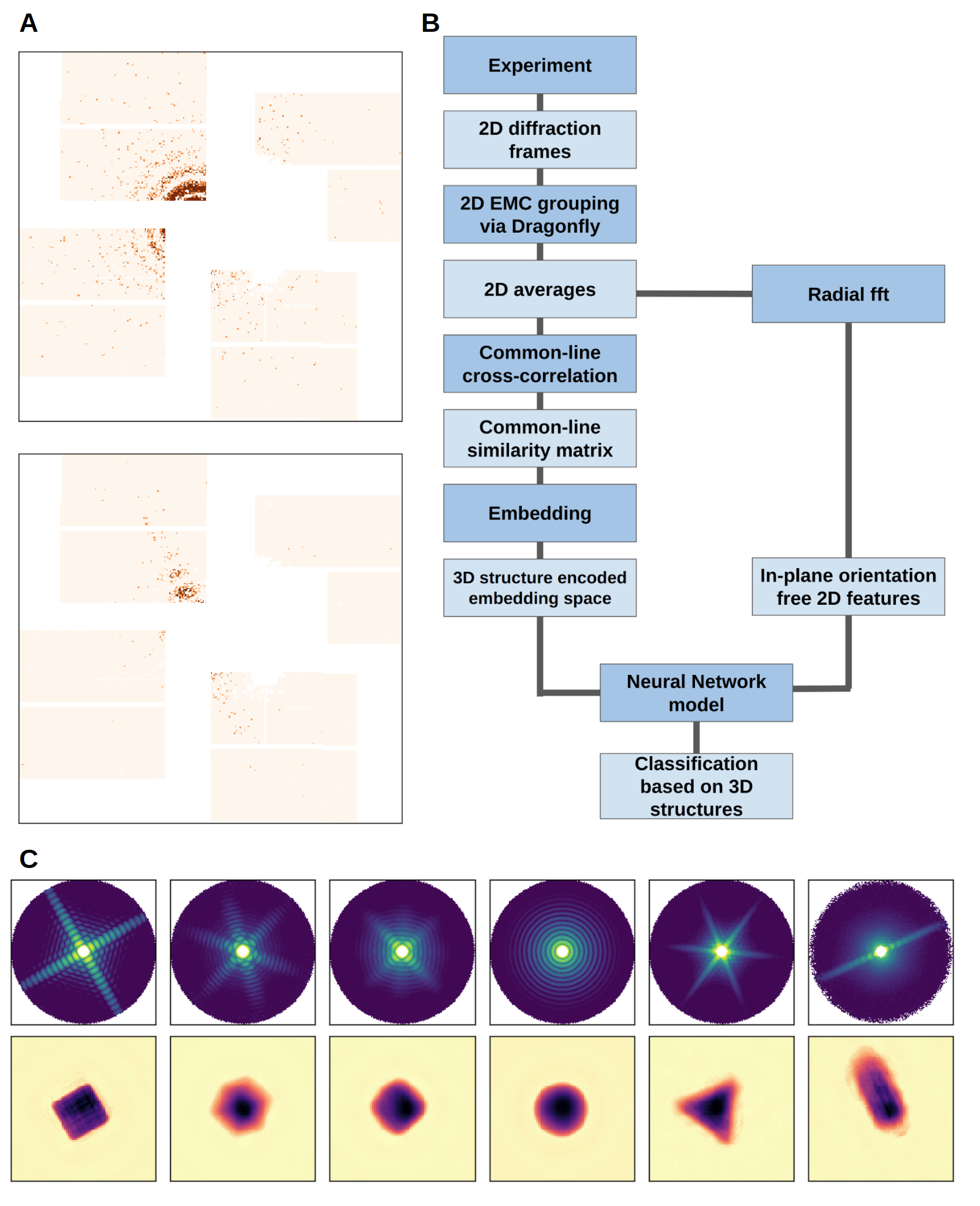}
\caption{(A) Examples of diffraction patterns in the data set. The color scale maximizes at 4 photons. (B) Workflow of the CLPCA method. (C) Upper row shows some typical frame averages on logarithmic scale in the dataset, while the lower row shows their corresponding real space 2D density projections via phase retrieval. The real-space field of view is 132.4 nm.}
\label{fig:frames_and_flow}
\end{figure}  

\subsection{Classification of the entire ensemble}
There are two major challenges in understanding the sample ensemble from the diffraction data. First, many of the patterns are too weak ($10^{2-3}$ photons/pattern) to apply single shot analysis, as illustrated in Fig.~\ref{fig:frames_and_flow}(A). Second, the dataset is not structurally homogeneous, not just in that the sample set includes non-cubic particles, but also that a fraction of the cubic particles suffered varied levels of melting/softening due to pre-exposure by the previous pulse in the train.

We first developed a common-line principal component analysis (CLPCA) method to understand SPI datasets from an arbitrary ensemble of particles. Fig.~\ref{fig:frames_and_flow}(B) shows our workflow, which we discuss in detail below. 
 
\subsubsection{\label{sec:dragonfly2d} Generation of 2D averages via bootstrapping}
Due to the weak scattering signal in single diffraction frames, the first step is to average similar 2D detector frames together (accounting for in-plane rotations), which was done with the 2D classification procedure implemented in \emph{Dragonfly}~\cite{Ayyer:2016}. The code implements a modification of the EMC algorithm that classifies all 2D frames into a given number of averages (models, or classes in Dragonfly). This averaging improves the signal-to-noise ratio and fills in detector panel gaps, at the potential cost of washing out some structural variations. In order to mitigate the latter effect, one can use a very large number of models, but this can lead to instabilities in the iterative reconstruction and reduced signal-to-noise in individual class averages. 

In order to get more pattern averages, a bootstrapping method was used by running the reconstruction 5 times with 200 models, each time with a random subset of 80\% of the frames. In this way, each of the 1000 models are composed of a different group of similar 2D frames. In Fig.~\ref{fig:frames_and_flow}(C), we show an example of some typical 2D diffraction averages and corresponding projected electron density maps after phase retrieval using a combination of the difference map and error reduction algorithms similar to that employed in \citeasnoun{Ayyer:2019}. This highlights the variety both in the samples, but also in the diffraction patterns from the same samples in different orientations, such as the rotated versions of identical cubes in the first two columns of Fig.~\ref{fig:frames_and_flow}(C).

\subsubsection{Common line 3D classification}
The 2D EMC method can only group similar 2D frames together, but as mentioned above, diffraction patterns of the same object can look very different depending on the orientation. In order to understand the variations of structures in the sample, we need to further classify the averages through their 3D features, rather than considering them just as 2D images.

\begin{figure}
\includegraphics[width=0.8\textwidth]{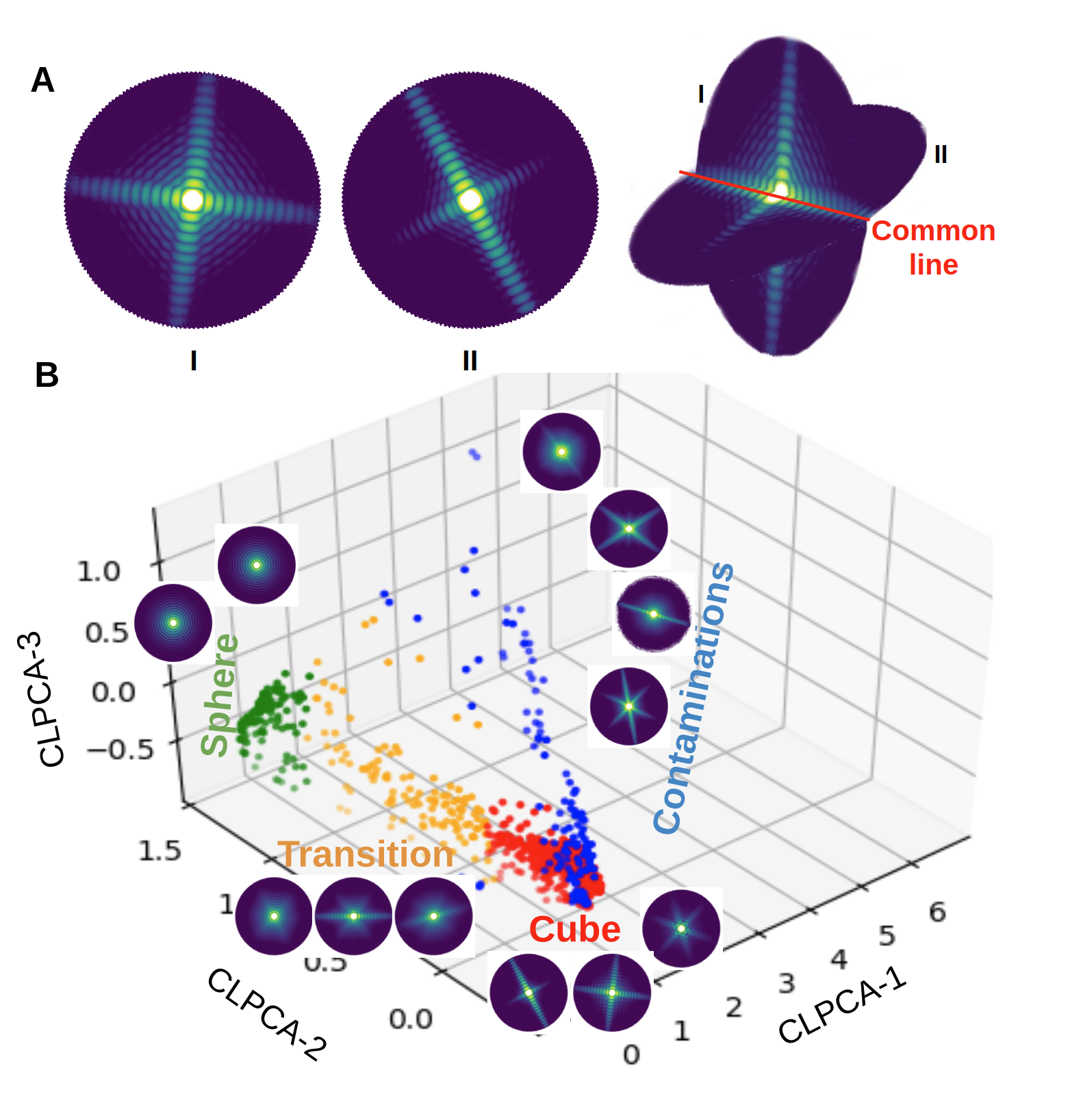}
\caption{(A) An illustration of the common-line between two patterns of similar shaped objects but different orientations. (B) The distribution of 1000 averages in the 3D CLPCA space. Different colors are manually divided groups from the first two components: contaminants (blue); cubes (red); transition (yellow); spheres (green). Typical patterns for averages in each group are also shown on logarithmic scale.}
\label{fig:PCA_space_overview}
\end{figure}

At small scattering angles, each average is a Fourier transform of a projection of the target object at a given orientation. According to the Fourier slice theorem, this means that each average represents a slice through the 3D Fourier transform of the object. Any two patterns of different orientations from the same object should share a common intersection line (at larger angles, these lines become arcs), as shown in the illustration of Fig.~\ref{fig:PCA_space_overview}(A). The similarity of the diffraction intensities along the best ``common-lines'' between two patterns should be correlated to the similarity of their 3D structures.  For each pair of 2D averages we defined the similarity between their common lines as the cross-correlation coefficient of the respective diffraction signals along those lines. The angles of the common lines are chosen to be those that yield the greatest degree
of similarity. This yields a common line similarity matrix (CC-matrix) for all pairs of 2D intensity averages. Common lines have been used previously for orientation determination~\cite{Shneerson2008,Singer:2011,Yefanov:2013}, where the optimal angles yell one how to fit the two slices in the 3D Fourier space. Here, we are interested in the similarity index of the common lines, as a tool for structural similarity analysis, rather than the relative angles at which they are maximized.

To get the landscape of the 3D structures in the sample, we performed dimensionality reduction on this CC-matrix and generate an embedding. Fig.~\ref{fig:PCA_space_overview}(B) shows a 3D embedding of the 1000 averages calculated using principal component analysis. In the rest of this article, we refer to this 3D embedding as `CLPCA space'. Until this point, the entire procedure is fully unsupervised and can be run automatically for any SPI dataset. 

By comparing with 2D patterns and their locations in CLPCA space, we then qualitatively divided the CLPCA space into 4 groups as shown in Fig.~\ref{fig:PCA_space_overview}(B), red - cubes, green - spheres, blue - contaminants and yellow - rounded cubes. 
In the embedded space, CLPCA-1 separates these assorted patterns from those originating from spherical/cubic particles, CLPCA-2 tracks the transition between spherical and cubic particles and for spherical particles, CLPCA-3 is associated with their diameter. In addition, we see a sequence where particles transition from cubes to spheres. This is a clear trace of the pre-melting processes observed in the experiment. Without using any \emph{a priori} knowledge about the sample set, we are able to obtain a rough classification of the dataset according to their 3D structures. 

We also note from Fig.~\ref{fig:PCA_space_overview}(B) that the dense cluster of cube patterns are correctly identified as being from the same shaped particles even though the patterns themselves vary extensively at different orientations. For the sake of simplicity we only perform the embedding with the PCA method, but other dimensionality reduction methods could also be used interchangeably. In Appendix~\ref{app:embeddings}, we show a comparison between PCA and other embedding methods, showing no strong preference in terms of separating structural classes for this dataset.

\subsubsection{Absolute Embedding of Images}
\label{sec:absolute}
The CLPCA method provides a way to classify unknown sample frames according to their 3D structures. But the `similarity' used for generating the landscape is a relative concept i.e. the distributions in the embedded space depend on the sample one chooses. This limits the classification and comparison of certain sub-samples. For example, one might want to look into some subgroup in detail while still be able to relate them to the whole-sample embedding, or generate new sets of averages with bootstrapping. In addition, the time complexity of calculating the similarity matrix scales as the square of number of models, which imposes a computational hurdle to using too many models at once.

\begin{figure}
\includegraphics[width=0.8\textwidth]{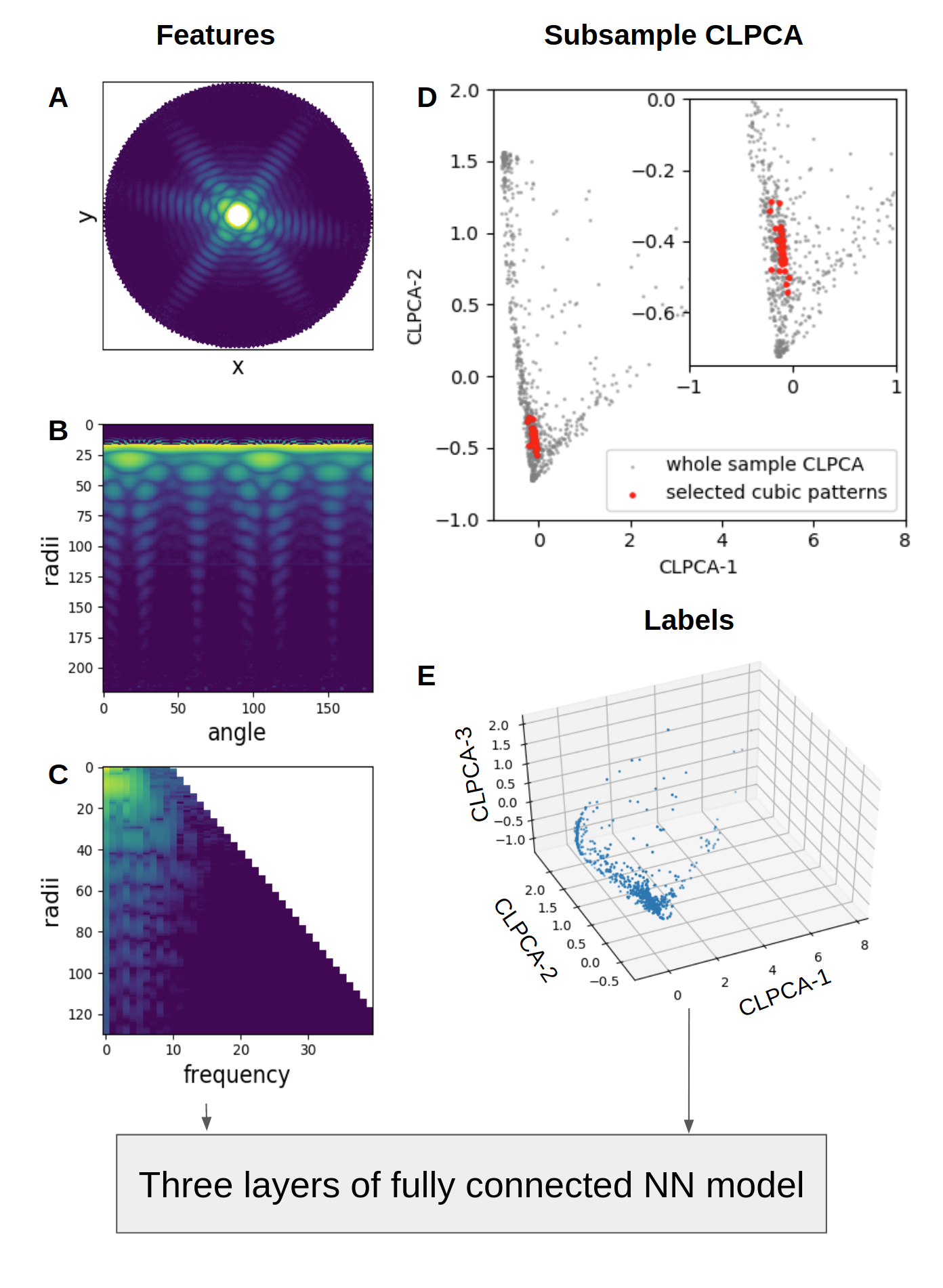}
\caption{Brief illustration of the absolute embedding neural network (NN) model. (A) Average pattern from \emph{Dragonfly}. (B) Polar representation of pattern. (C) Stack of 1D Fourier transform magnitudes along the angular axis for each radial bin. The odd frequency components (due to inversion symmetry) as well as the higher frequencies for signals at smaller radii are removed. These represent the feature vectors for the neural network. (E) Training labels from CLPCA. (D) Example of using absolute CLPCA on a selected cubic subset of frames (red dots). The gray dots represent the embedding of the pattern averages from the whole dataset.}
\label{fig:NN_FL}
\end{figure}

In order to get a quicker and more universal measure of frame features, once we obtain the embedding of a typical set of averages, a neural network-based regression was used to map any previously unseen averages into the defined embedding space. Firstly, we extracted the relevant features of the patterns: for each of our 1000 averages we Fourier-transformed its azimuthal intensity variation at every radius and kept the absolute values to make the pattern invariant to in-plane rotations. We then kept frequency signals within the spatial resolution at each radius as the training features, as shown in Fig.~\ref{fig:NN_FL}(A-C). We used the coordinates of the averages in CLPCA space as training labels, as shown in Fig.~\ref{fig:NN_FL}(E). 

The neural network used for fitting the relation between the pattern of 2D averages and their coordinates in CLPCA space has four fully connected hidden layers with 512, 128, 64, 32 nodes per layer. From the 1000 patterns used to calculate the similarity matrix, 800 were used to train the model, which was then validated with the rest of the 200 patterns with mean square error of 0.088, 0.059 and 0.042 for components 1, 2 and 3. With this model, we were able to quickly find the absolute embedding of any single 2D intensity model and zoom into arbitrary subsets of patterns while still retaining a reference to the full data set. An example is shown in Fig.~\ref{fig:NN_FL}(D) where we use CLPCA on a manually selected subset of averages in the ``cube'' region of CLPCA space. The frames which contributed to these selected averages were classified using \emph{Dragonfly}. The embedding of these new averages is shown in red with reference to the whole-dataset CLPCA embedding.

In summary, the CLPCA method provides a robust and mostly parameter-free framework to classify and visualize the structural landscape of an arbitrary set of coherent diffraction patterns. 

\subsection{3D reconstruction of heterogeneous models}
In the previous section we found a sequence where the shape of the particles transitioned from cubic to spherical. To further understand this transition, we would like to be able to reconstruct the 3D models along the sequence. Given a set of diffraction patterns from a discrete set of reproducible objects, it is relatively straightforward to generate multiple 3D models using the EMC algorithm~\cite{Ayyer:2021,Cho2021}. Here, this approach is difficult due to two reasons, 1.) it needs to assume a number of discrete heterogeneity models, which does not qualitatively capture the continuous cube-sphere transition and 2.) we do not have enough patterns located in the transition region to reconstruct 3D structures via EMC.

Fortunately, previous studies in Cryo-EM single-particle analysis provide us a good way of modeling the continuous heterogeneity. \citeasnoun{Zhong:2021} developed cryoDRGN, a variational auto-encoder (VAE) to efficiently reconstruct heterogeneous complexes and continuous trajectories of protein motions~\cite{Kingma:2019}. Inspired by their paper, we developed a similar deep learning model by combining a VAE and convolutional neural networks (CNNs) to model the continuous shape transition along this ``melting'' sequence. 

\subsubsection{Variational Auto-encoders}
\begin{figure}
\includegraphics[width=0.9\textwidth]{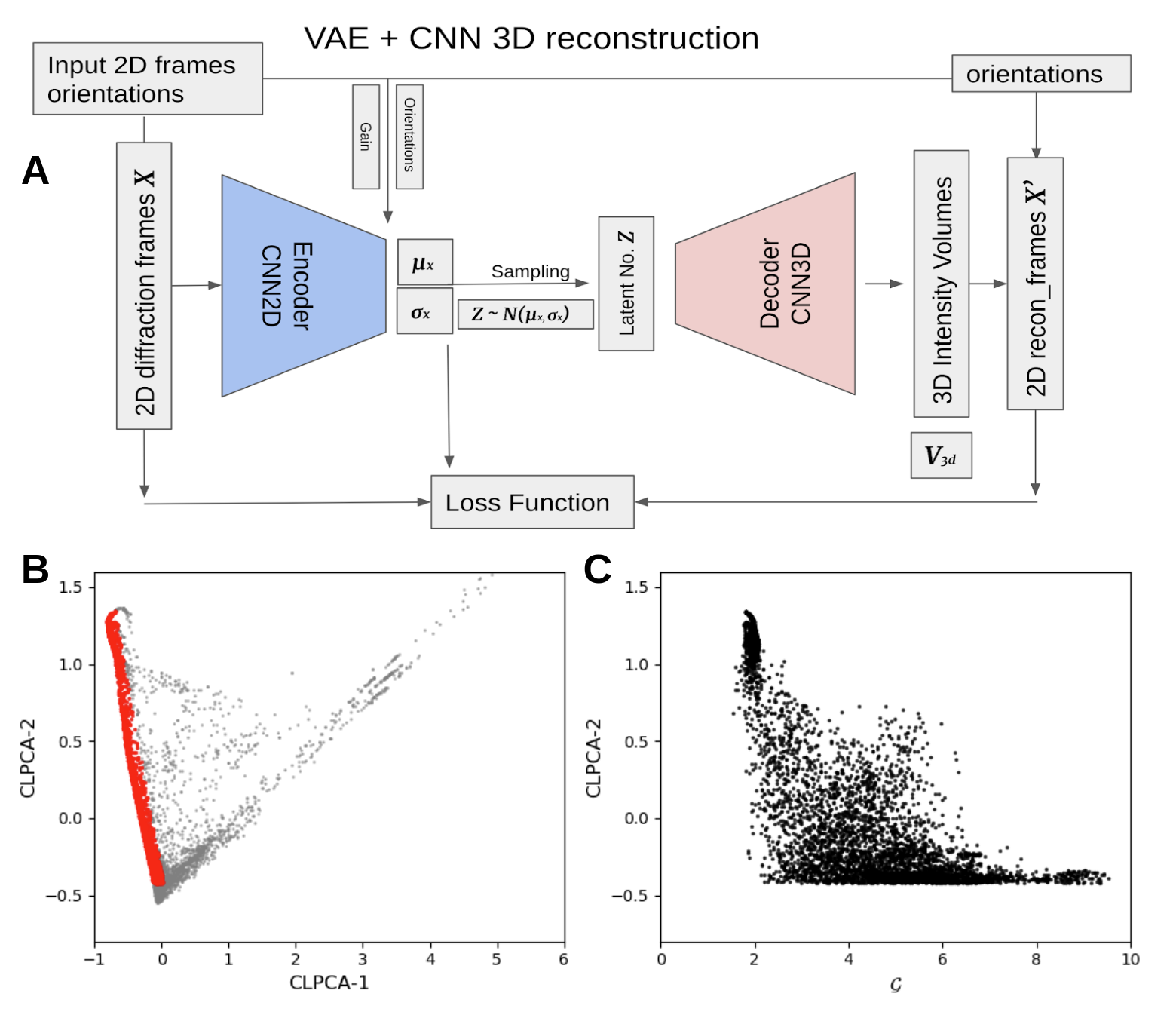}
\caption{(A) Schematic of the variational auto-encoder (VAE). It consists of two convolutional neural networks as encoder and decoder, and the model is fit by comparing the similarity between input and decoder generated 2D slices with additional regularization by assuming the latent parameter follows a $N(0,1)$ distribution. (B) The distribution of the bootstrapped 10000 2D average patterns in CLPCA space (gray). Red dots are selected average patterns along the melting sequence used for the VAE analysis. (C) CLPCA-2 against $gain$ for the selected patterns from (B) along the melting sequence.}
\label{fig:VAE_CNN_scheme}
\end{figure}

Figure~\ref{fig:VAE_CNN_scheme}(A) shows the architecture of the VAE neural network, consisting of a 2D CNN pattern-encoder to encode 2D patterns along with their orientation estimates into distributions of latent parameters, and a 3D transposed convolution network as volume-decoder to generate 3D intensity volumes from latent numbers. This setting allows the neural networks to learn the 3D-heterogeneity-structure-encoded latent numbers from the diffraction patterns.

The inputs of the VAE network are the 2D intensity averages used as input for the CLPCA method in the previous section along with their associated relative orientations. The former are obtained from the \textit{Dragonfly} output discussed in Section~\ref{sec:dragonfly2d}, while the latter are calculated as described below. We start with the 3D intensity volume of an ideal 42-nm cubic particle, slicing it with 16407 orientations uniformly distributed in quaternion space within the $O_h$ sub-group to account for the symmetry of the object. The cross-correlation coefficient (CC) of each 2D model is calculated with all the orientations. The orientation with the highest CC is recorded for each model. In addition, another parameter which helps in training is termed \emph{gain} referring to the ratio of the best to the average CC over all orientations. This parameter can also indicate the 'cubicness' of the particle since patterns from cubes fit very well to the synthetic model. 

The VAE works in the following manner: For each input average $X$, its orientation, $\Omega$ and gain, $\mathcal{G}$, the encoder network generates a Gaussian distribution of its latent variable values $N(\mu, \sigma)$. The fact that this is not just a single latent vector $z$, but a distribution makes the auto-encoder variational and enforces the smoothness of the latent space. It also enables the network to use information from neighbouring regions in the space to update regions with limited data. A latent vector $z$ is sampled using this distribution and used by the decoder to generate a 3D intensity volume $V_{3d}$, which is then symmetrized by the octahedral point group. The known orientation is then used to slice the generated 3D volume to get a reconstructed 2D pattern $X'$. The goal of training is to minimize the difference between $X$ and $X'$ (further details in Appendix~\ref{app:vae_details}). 3D information is obtained since the same latent space region is sampled by averages in a variety of orientations. Note, that although the $\mathcal{G}$ is not strictly necessary for the VAE network to reconstruction 3D volumes, it helps to regularize the latent space of $z$.

When the model is trained, we can encode the 2D diffraction patterns into the latent space via the encoder network. The VAE network not only reconstructs the 3D structure of any given input 2D diffraction pattern but can also do so for any chosen location in the latent space. In this way, it can be used to study the continuous transition between cube and sphere in the melting sequence, including in regions where there aren't sufficient patterns to isolate and obtain a conventional reconstruction. 

\subsubsection{Tracing the melting sequence}
As mentioned before, using \emph{Dragonfly} to generate 2D averages comes at the cost of potentially averaging out structural variations in individual frames due to the limited number of classes. To generate the CLPCA landscape, 1000 averages were deemed to be enough to cover most of major shapes in the sample, but in order to trace the continuous shape variation along the melting sequence more detailed minor variation information needs to be retained. To keep more of those variation information along the melting sequence we repeat the same bootstrapping plus \emph{Dragonfly} method described in Section~\ref{sec:dragonfly2d} to generate 10,000 average patterns. 

Using the CLPCA method in combination with the absolute embedding approach discussed in Section~\ref{sec:absolute}, we selected 5965 of these averages located along the melting sequence. Fig.~\ref{fig:VAE_CNN_scheme}(B) shows the distribution of these averages and the selected melting sequence averages in CLPCA space. As discussed above, the CLPCA-2 is roughly aligned with the melting sequence, which can be used as a coarse estimator of the melting process. Fig.~\ref{fig:VAE_CNN_scheme}(C) shows the comparison between CLPCA-2 and the gain $\mathcal{G}$ for the selected melting sequence models. Although correlated with the melting sequence at a certain level, by visual inspection $\mathcal{G}$ seems to be good at tracing the cubes while CLPCA-2 is better in isolating the spheres. 

In order to reduce computational cost, the input 2D average intensities were downsampled and slightly truncated at high $q$. Since the intensity distribution of a compact object is heavily weighted to low $q$, the input data was normalized by the azimuthally averaged intensity over the whole dataset. Details of the VAE network structure as well as various pre-processing steps are given in Appendix~\ref{app:vae_details}.

\begin{figure}
\includegraphics[width=0.75\textwidth]{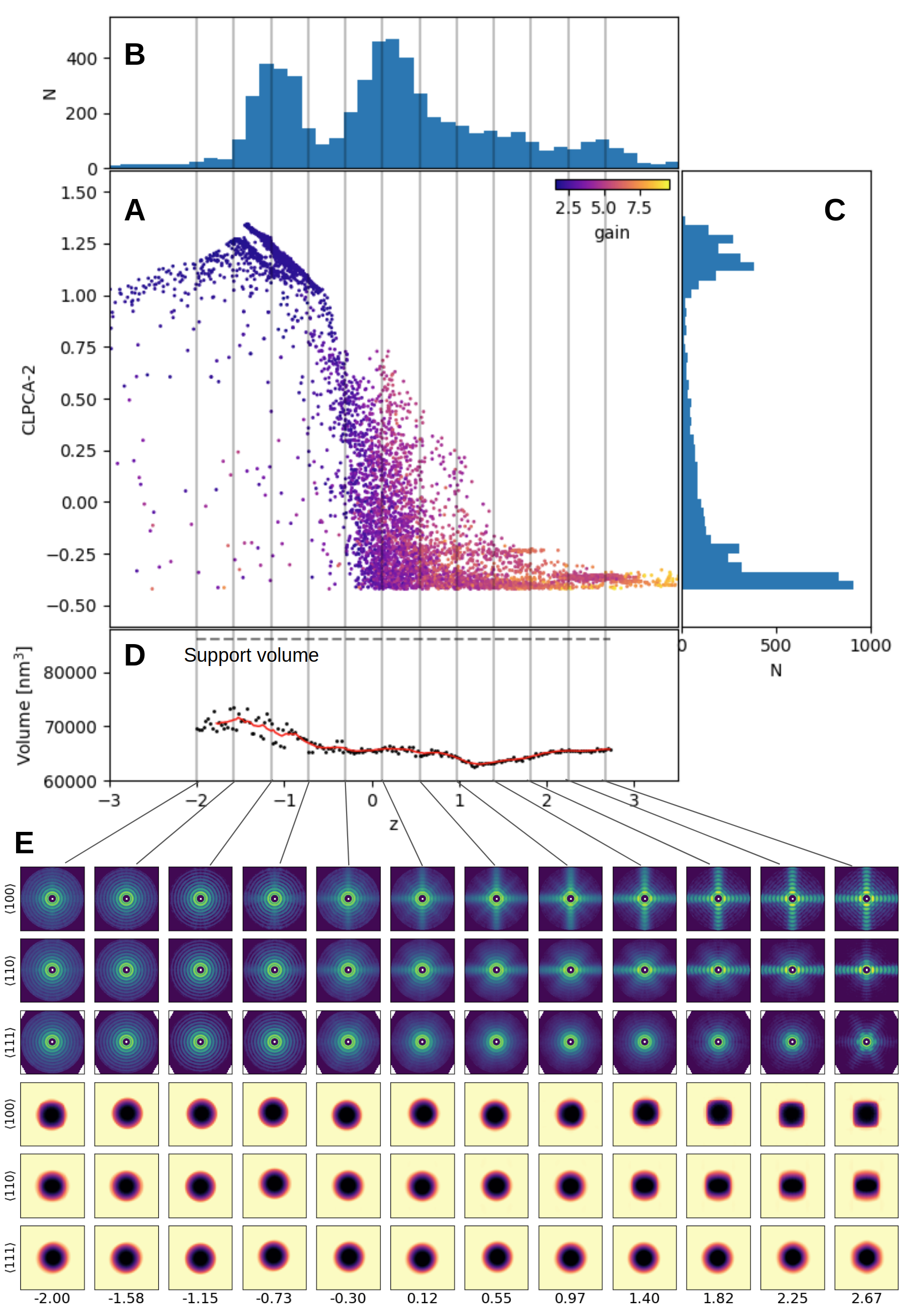}
\caption{(A) VAE encoded 1D latent number $z$ against CLPCA-2 for the input sample patterns, color coding is the $gain$ parameter of each pattern. (B) Histogram for latent number $z$. (C) Histogram for CLPCA-2. (D) Volume ``evolution'' along $z$, volumes calculated with density threshold equal to $10^{-4}$ total mass. Dashed horizontal line is the size of the support volume in phase retrieval. Vertical gray lines shows the location of 12 selected $z$ numbers. (E) Top three rows are the VAE-generated intensity volumes from the 12 selected $z$ numbers on logarithmic scale, three rows shows slices in (from row 1 to row 3) $\langle100\rangle$, $\langle110\rangle$ and $\langle111\rangle$ directions respectively. Bottom three rows show their corresponding density projections in real space.}
\label{fig:VAE_gain_z1}
\end{figure}  

For simplicity, we start by modeling with a one-dimensional latent space, so the VAE model is forced to trace the strongest variation along the melting sequence/cubic-spherical transition. Fig.~\ref{fig:VAE_gain_z1}(A) shows the VAE encoded 1D latent parameter $z$ against CLPCA-2 for the input models, color coded by the $\mathcal{G}$ parameter: brighter yellow indicates the particles are more anisotropic. Fig.~\ref{fig:VAE_gain_z1}(B) and (C) are the histograms of $z$ and CLPCA-2 respectively. Though there exists a strong correlation between latent parameter $z$ and CLPCA-2, they show different distributions in tracing the cubic--spherical transition. First, the VAE further separates the cubic side cluster classified by CLPCA (CLPCA-2 $> 0.5$). Secondly, at lower than $z\approx -1.5$ another sequence is visible which is not clearly seen in CLPCA space.  

One advantage of the VAE network is that it allows us to reconstruct/generate the 3D intensity volume for any given $z$. Here, we selected 12 $z$ numbers of equal steps between -2.00 and +2.67. In Fig.~\ref{fig:VAE_gain_z1}(E), the upper three rows show three special ($\langle100\rangle$, $\langle110\rangle$ and $\langle111\rangle$) slices of the VAE-generated 12 intensity volumes on logarithmic scale. We see a clear and smooth cubic-spherical transition from $z\approx 2.7 \rightarrow -1.5$, which shows that the VAE network is able to trace the early melting stages much better than the CLPCA. 
And for the second `sequence' at $z < 1.5$, the lower contrast in the intensities and less spherical structures indicate they are likely to be the contaminated particles at late melting stages, the low contrast being due to the fact that they contain various shapes and lack of accurate orientation estimates. Compared with the CLPCA method that roughly classifies cubic and spherical particles, the VAE is able to provide a detailed, smooth modeling of the entire transition process. 

The bottom three rows of Fig.~\ref{fig:VAE_gain_z1}(E) show projections of phase-retrieved density maps in real space along the same three directions. We also find the volume increases by around 10\% from cubes to spheres along the melting sequence by calculating the volume evolution of 3D reconstructed volumes at finely sampled 200 different $z$-values as black dots shown in Fig.~\ref{fig:VAE_gain_z1}(D). This is consistent with the density difference between crystalline gold (19.32 g/cc) and molten/randomly close packed atoms (17.31 g/cc). This is also in agreement with the direct size fitting on 2D average patterns of spherical and cubic particles, shown in the Fig.~\ref{fig:volume_comparison}.

This volume analysis allows us to observe an additional feature, namely that the melting sequence seems to show two different stages: Stage 1 from $z\approx 3 \rightarrow -0.3$, where most of the shape change occurs, but with no significant volume changes. Stage 2 from $z\approx -0.3 \rightarrow -1.5$ where all particles are mostly round but the volume increases. This is to be expected since edges and vertices require lower energies to be disrupted than for complete melting, where the bulk lattice structure is lost~\cite{Cahn:1986,Chen:2021}. As discussed below, the melting sequence here is only based on the 3D structure of particles. Thus, the “two stages” could only suggest that the shape changing happened ``before'' the size expansion along $z$ (exposure intensity). 

We note that the melting sequence here is purely based on the 3D structure of particles. Since we observe a continuous transition, we assume a monotonic relationship between the latent variable $z$ and the incident fluence in the previous pulse. However, without additional information about the expected distribution of particles with a given incident fluence, we cannot obtain the precise mapping between the two. As all pre-exposures happened 880 ns before the imaging pulse, one should not view the melting sequence as a time-dependent shape variation, but rather an incident fluence- or temperature-dependent behavior.

\begin{figure}
\includegraphics[width=0.99\textwidth]{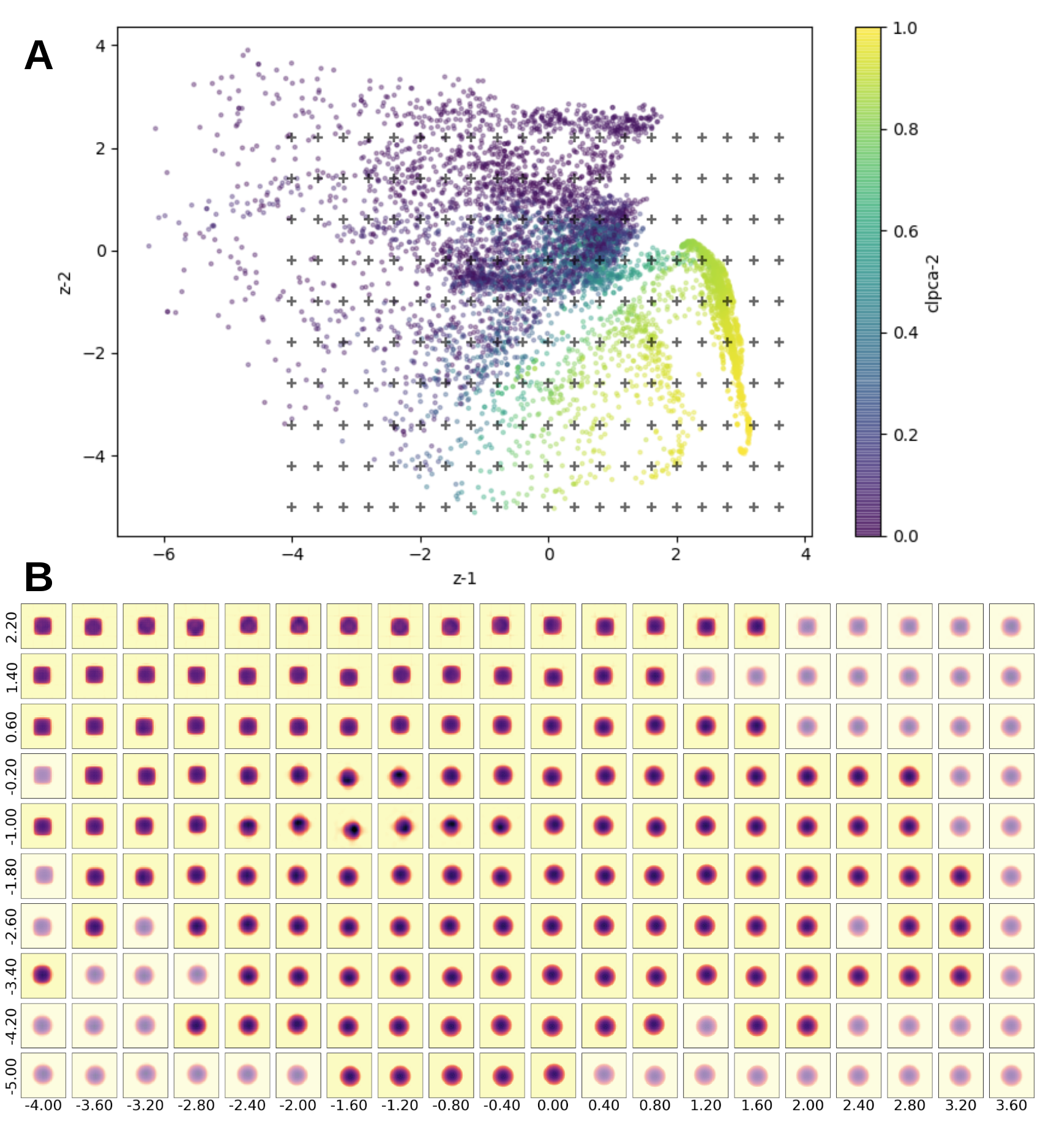}
\caption{(A) The distribution of the dataset in 2D latent space, color coded by CLPCA-2. Black crosses in the plot are select 200 latent numbers $z$. (B) The $\langle100\rangle$ direction projection of 200 real space density retrieved from the logarithmic intensity volumes generated from the 200 selected latent numbers $z$ (black crosses in (A)). Transparent slices are volumes generated from latent regions without data points.}
\label{fig:Latent_2D}
\end{figure}  

We have shown that the one dimensional latent space VAE network is capable of modeling the major transition, but it is not the only variation in the dataset. To trace more variations we now model it with a higher dimensional latent space. Fig.~\ref{fig:Latent_2D}(A) shows the VAE-encoded 2D latent parameters of the sample. Neither the first latent nor the second latent parameter traces the cubic - spherical transition independently. This shows the encoding of more secondary variations, e.g. contaminant, size etc.  

We selected 200 evenly distributed points in the latent space and generate their 3D intensity volumes. Fig.~\ref{fig:Latent_2D}(B) shows the $\langle100\rangle$ direction projection maps of the phase retrieved VAE generated intensity volumes of the 200 selected latent numbers $z$. One can clearly notice the transition between spherical to cubic features from right to the left is roughly encoded in the first $z$ component. However, the trend is twisted in 2D space, we got multiple transition sequences along the second component. In regions with no or very few data points (half transparent slices), the results are less reliable as the VAE network could not generate reliable intensity volumes.  
 
\section{Conclusion}

In this work, we demonstrate two methods to study heterogeneous ensembles of samples using X-ray single particle imaging by analyzing the relationships between the 3D structures of the samples rather than the diffraction patterns directly. The first, termed CLPCA, uses the fact that diffraction patterns from ideal particles share a common line/arc from the Ewald construction. The similarity of the best common lines allows us to visualize a low-dimensional embedding of the dataset which roughly represents 3D similarity regardless of orientation.  A fully-connected neural network was trained against the embedding in order to embed arbitrary 2D averages with respect to the rest of the data set.

This was applied to an experimental dataset from gold nanoparticles, containing 2.4 million patterns with only a few thousand photons per frame. In order to improve the signal-to-noise ratio and fill in detector panel gaps, 2D classification in \emph{Dragonfly} was used to combine similar patterns up to in-plane rotations, and a bootstrapping approach was used to increase the number of 2D averages with subsets of the data.

The second method involves a generative VAE network that models continuous structural deformations, and can be used to generate 3D structures at arbitrary points along the landscape. In the studied dataset, this network was able to recover a continuous melting sequence induced by pre-exposure of the particles by the previous pulse in the European XFEL pulse train. We observe an initial rounding of the particles followed by a slight expansion caused by atomic-scale disordering at higher temperatures. Our two methods open up the possibility of studying samples of heterogeneous particles, and potentially tracing dynamic motions of particles in the SPI experiments.

\appendix
\section{Different embeddings of CC-matrix}
\label{app:embeddings}

\begin{figure}
\includegraphics[width=0.75\textwidth]{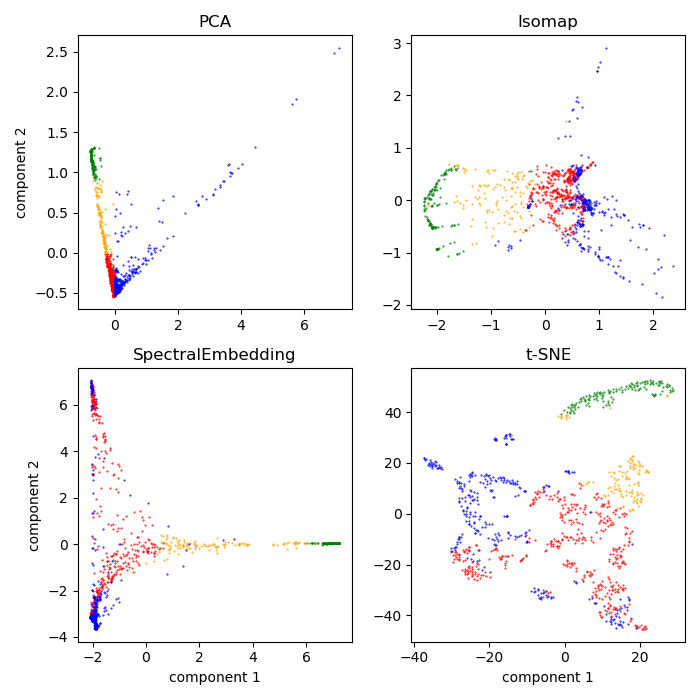}
\caption{Comparison between four different common embedding methods, top left: PCA, top right: Isomap, bottom left: Spectral, and bottom right: t-SNE. Color coding is same as in Fig.~\ref{fig:PCA_space_overview} by manually divided groups from the first two components of PCA space.}
\label{fig:3_embeddings}
\end{figure}

In Fig.~\ref{fig:3_embeddings} we show a comparison between PCA and other standard embedding methods implemented in the \texttt{scikit-learn} Python package~\cite{Pedregosa:2011}. For the sample set considered, all four embeddings are able to separate the 4 basic shape groups (cube, sphere, transition and contaminant), although this may not be true for other, more complex continuous structural variations.

\section{Details of the VAE model}
\label{app:vae_details}
\subsection{Preprocessing the input patterns}
The 2D averages from \emph{Dragonfly} have an initial size  of $441 \times 441$ pixels which is computationally redundant considering the data is highly oversampled. For the results shown in this paper, we used the following procedures to reduce the size. We firstly downsampled the size to $243\times243$ and then we cut off the high $q$ part to further reduce the size to $161\times161$.
 
In addition to the size reduction, diffraction patterns of compactly supported objects are overwhelmingly dominated by the low $q$ signal. To appropriately weight the higher $q$ shape information, we divided the 2D intensities by the radial average intensity over the whole dataset before feeding them into the VAE and multiplied it back when generating the 3D intensities. 

\subsection{Model parameters}
The depth and nodes of our encoder/decoder network depend on the quality of the sample and the spatial resolution of reconstructed 3D volumes we need. In our code we provide 3 different reconstructing volume sizes options: low volume ($81^3$), intermediate volume ($161^3$) and high volume ($243^3$). For the results shown in this paper, the intermediate volume model is used in which the encoder network has four 2D convolution layers with ($16 \times 81^2$), ($32 \times 27^2$), ($64 \times 9^2$) and ($128 \times 9^2$) nodes and ($3 \times 3$) kernel size per each layer. The decoder consists of six 3D transposed convolution layers with ($128 \times 9^3$), ($64 \times 9^3$), ($32 \times 27^3$), ($16 \times 81^3$), ($8 \times 161^3$) and ($1 \times 161^3$) nodes and ($3 \times 3 \times 3$) kernel size per each layer.

\begin{figure}
\includegraphics[width=0.5\textwidth]{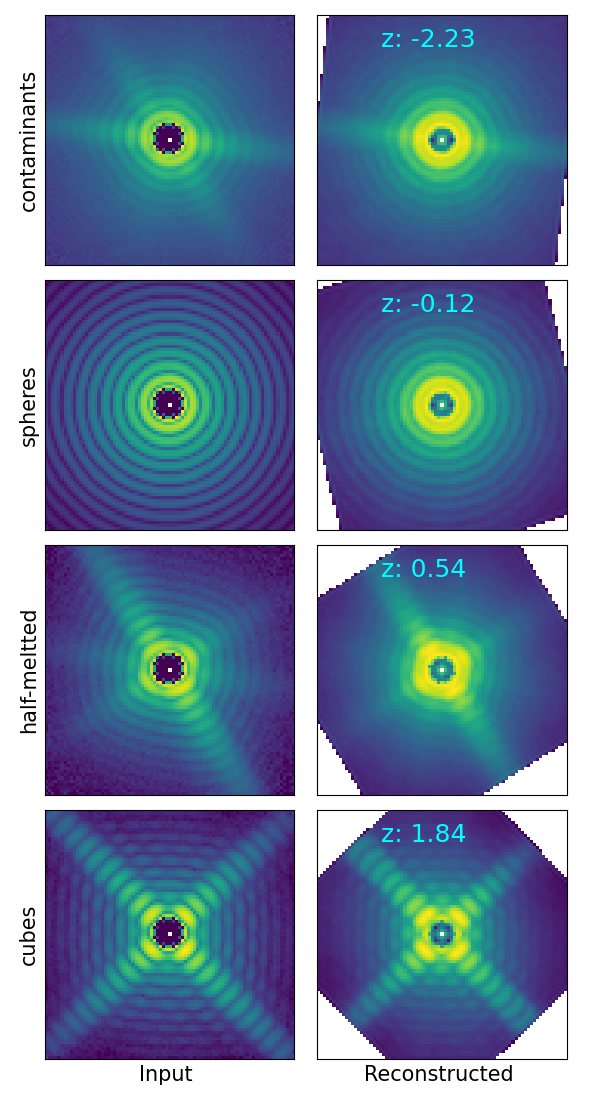}
\caption{Comparison between input intensities and reconstructed intensity patterns on logarithmic scale.}
\label{fig:Sanity_check}
\end{figure}

In Fig.~\ref{fig:Sanity_check} we show a comparison of input 2D patterns and their corresponding 2D slices of their reconstructed 3D intensities at four representative latent z-values from the model shown in Section 2.3.2. Together with Fig.~\ref{fig:VAE_gain_z1} the first pattern is likely from half-melted contaminant particles, where the reconstruction is not reliable because the number of patterns of that shape are low and the orientation estimate by fitting against a cube is inaccurate. The other rows are well reconstructed spheres, melting cubes and cubes.

\subsection{Devices}
The network was implemented in PyTorch and run on a single node with 4 NVidia V100 GPUs with 32 GB memory each. For our 5672 2D averages, the $161^3$ volume model takes less than 2 hours to calculate.

\section{Direct size comparison of cubes and spheres}

\begin{figure}
\includegraphics[width=0.5\textwidth]{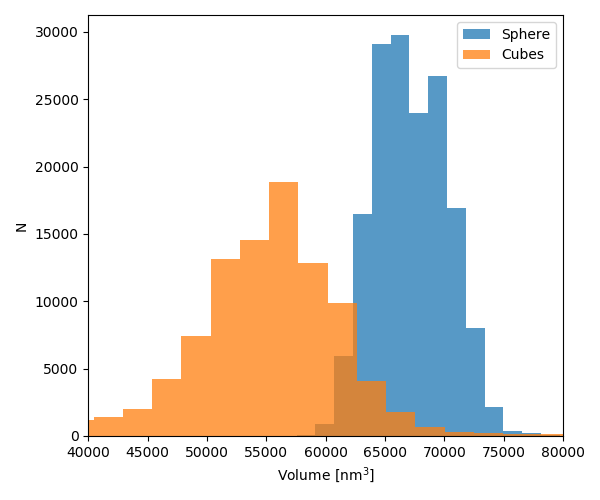}
\caption{Volume distribution of cubic particles (orange) and spherical particles (blue).}
\label{fig:volume_comparison}
\end{figure}  

Figure~\ref{fig:volume_comparison} shows the volume comparison of classified pure cubes and pure spheres. The cube sizes were calculated directly from frames with clear streak fringes, which were assumed to have at least one set of faces parallel to the X-ray beam. The larger spread of cube sizes can be accounted for by the inclusion of frames where the faces were not exactly parallel to the beam. Nevertheless, similar to the VAE model, a clear size increase is observed from cubes to spheres.




\ack{We acknowledge European XFEL in Schenefeld, Germany, for provision of X-ray free-electron laser beamtime at Scientific Instrument SPB/SFX (Single Particles, Clusters, and Biomolecules and Serial Femtosecond Crystallography) and would like to thank the staff for their assistance. This work has been supported by the Clusters of Excellence ‘Center for Ultrafast Imaging’ (CUI,EXC 1074, ID 194651731) and ‘Advanced Imaging of Matter’ (AIM,EXC 2056, ID 390715994) of the Deutsche Forschungsgemeinschaft (DFG). This work has also been supported by the European Research Council under the European Union’s Seventh Framework Programme (FP7/2007-2013) through the Consolidator Grant COMOTION (ERC-614507-Küpper) and by the Helmholtz Gemeinschaft through the “Impuls-und Vernetzungsfond”. J.C.H.S. and R.A.K. acknowledge support from the National Science Foundation BioXFEL award (STC-1231306). F.R.N.C.M. acknowledges support from the Swedish Research Council, Röntgen-Ångström Cluster and Carl Tryggers Foundation for Scientific Research. P.L.X. acknowledges a fellowship from the Joachim Herz Stiftung. P.L.X. and H.N.C. acknowledge support from the Human Frontiers Science Program (RGP0010/2017).}

\referencelist[main]



\end{document}